\documentclass[prl,preprintnumbers,twocolumn,eqsecnum,floatfix,nofootinbib,a4paper,superscriptaddress]{revtex4-1}
\usepackage{color}
\usepackage{calc}
\usepackage{amsmath,amssymb,graphicx}
\usepackage{amssymb,amsmath}
\usepackage{tensor}
\usepackage{bm}
\usepackage{float}
\usepackage{microtype}
\usepackage{booktabs}
\usepackage{times}
\usepackage[varg]{txfonts}
\usepackage[colorlinks, pdfborder={0 0 0}]{hyperref}
\usepackage{subfigure}
\definecolor{LinkColor}{rgb}{0.75, 0, 0}
\definecolor{CiteColor}{rgb}{0, 0.5, 0.5}
\definecolor{UrlColor}{rgb}{0, 0, 0.75}
\hypersetup{linkcolor=LinkColor}
\hypersetup{citecolor=CiteColor}
\hypersetup{urlcolor=UrlColor}
\allowdisplaybreaks
\usepackage[utf8]{inputenc}
\usepackage{ulem}
\usepackage{comment}
\usepackage{hyperref}
\usepackage[nolist]{acronym}
\usepackage{tikz}
\usepackage{textcomp}
\usepackage[export]{adjustbox}
\usepackage{url}

\newcommand{\Dl}{D_\ell}
\newcommand{\zl}{z_\ell}
\newcommand{\zs}{z_s}
\newcommand{\Ds}{D_s}
\newcommand{\Dls}{D_{\ell s}}
\newcommand{\vOmega}{\vec{\Omega}}

\usetikzlibrary{arrows,shapes,trees,decorations.pathreplacing}
\usepackage{aas_macros}

\usepackage{romannum}

\allowdisplaybreaks

\begin{document}

\title{Cosmography using strongly lensed gravitational waves from binary black holes}

\author{Souvik Jana}
\affiliation{International Centre for Theoretical Science, Tata Institute of Fundamental Research, Bangalore 560089, India}

\author{Shasvath J.  Kapadia}
\affiliation{International Centre for Theoretical Science, Tata Institute of Fundamental Research, Bangalore 560089, India}
\affiliation{The Inter-University Centre for Astronomy and Astrophysics, Post Bag 4, Ganeshkhind, Pune 411007, India}

\author{Tejaswi Venumadhav}
\affiliation{Department of Physics, University of California at Santa Barbara, Santa Barbara, CA 93106, USA}
\affiliation{International Centre for Theoretical Science, Tata Institute of Fundamental Research, Bangalore 560089, India}

\author{Parameswaran Ajith}
\affiliation{International Centre for Theoretical Science, Tata Institute of Fundamental Research, Bangalore 560089, India}
\affiliation{ Canadian Institute for Advanced Research, CIFAR Azrieli Global Scholar, MaRS Centre, West Tower, 661 University Ave, Toronto, ON M5G 1M1, Canada}

\begin{abstract}
Third generation gravitational wave (GW) detectors are expected to detect millions of binary black hole (BBH) mergers during their operation period. A small fraction of them ($\sim 1\%$) will be strongly lensed by intervening galaxies and clusters, producing multiple observable copies of the GW signals. The expected number of lensed events and the distribution of the time delay between lensed images depend on the cosmology. We develop a Bayesian analysis method for estimating cosmological parameters from the detected number of lensed events and their time delay distribution. The expected constraints are comparable to that obtained from other cosmological measurements, but probing a different redshift regime ($z \sim 10$) that is not explored by other probes. 
\end{abstract}

\maketitle

\section{Introduction}\label{sec:introduction}

The first gravitational wave (GW) detection by Advanced LIGO \citep{aLIGO} announced the arrival of a new astronomy\cite{GW150914}. Since then, $\sim$100 GW signals~\cite{gwtc-1, gwtc-2, gwtc-3, ias-1, ias-2, ias-3, ias-4, 1-ogc,2-ogc,3-ogc} have been detected by LIGO and Virgo \cite{Virgo}, the majority of which correspond to stellar-mass binary black hole (BBH) mergers. The remaining {detections correspond to} neutron star-black hole (NSBH) \cite{nsbh} and binary neutron star (BNS) mergers \cite{GW170817, GW190425}.  

The astrophysical riches extracted from the GW observations have been spectacular.  These have enabled unique tests of general relativity \cite{tgr-gwtc2, tgr-gwtc1, tgr-GW150914,GW170817-GRB170817}, provided the very first glimpse of the population properties of BBHs \cite{rp-gwtc3}, and ruled out a number of neutron star equation of state models \cite{gw170817-eos}. GWs have also given us the ability to estimate luminosity distances to their sources without using distance ladders. This in turn has provided new measurements of the Hubble constant $H_0$ \cite{LIGOScientific:2017adf,LIGOScientific:2019zcs,cosmo-gwtc3,DES:2019ccw,LIGOScientific:2018gmd,Palmese:2021mjm,Gray:2021sew,Mukherjee:2022afz}. While their current uncertainties are large, future detections will enable precise measurements (see, e.g., \cite{chen-hubble-3g, Farr:2019twy,You:2020wju,Ezquiaga:2022zkx,Shiralilou:2022urk} for prospective constraints, and \cite{Mastrogiovanni:2021wsd,Mukherjee:2021rtw,2022arXiv220403614H} for some caveats).

The proposed third generation (3G) network of ground-based detectors, consisting of two Cosmic Explorers \cite{CosmicExplorer} and one Einstein Telescope \cite{EinsteinTelescope}, is planned to have a markedly improved sensitivity --  by as much as an order of magnitude -- {compared to that of} the Advanced LIGO {and Virgo} detectors.  The expected detection rate in the 3G era is therefore enormous {($\sim 10^5 - 10^6$} BBH events per year \cite{ETScienceCase}). These detectors could observe stellar mass BBHs at redshifts of up to $z \sim 100$ \cite{HallEvans3G}.  

Using GW standard sirens, 3G detectors could not only enable a stringent constraint on $H_0$,  but also potentially provide precise measurements of other cosmological parameters, such as the matter density $\Omega_m$ and cosmological constant density $\Omega_{\Lambda}$  of the flat $\Lambda$CDM cosmological model\cite{chen-hubble-3g, Farr:2019twy,You:2020wju,Ezquiaga:2022zkx,Shiralilou:2022urk}.  While precise estimates of these parameters have been obtained from the study of the cosmic microwave background (CMB) \cite{Planck18}, Type \Romannum{1}a supernovae~\cite{Riess_2022}, etc, there is considerable value in providing independent constraints using GWs. Currently, there appears to be an inconsistency between the high-redshift ($z \sim 1000$) CMB data and the low-redshift  ($z\lesssim 2$) probes such as supernovae~\cite{Riess_2022}. This could could be a result of unknown systematic errors or point to the breakdown of the $\Lambda$CDM  model. Not only would these GW-based measurements  have different systematic errors, they would also probe the cosmology {using data from} redshifts ($z \sim 2-15 $), not probed by {the} CMB or other electromagnetic (EM) observations. 

Strong lensing of GWs offers a unique complementary probe of cosmology. Strong lensing can produce multiple observable copies of the GW signals which arrive at the detector at different times. During their {operational lifetime} ($\sim$10 years), 3G detectors are expected to observe tens of thousands of GW signals strongly lensed by galaxies and galaxy clusters. 

The most famous cosmological probe involving strong lensing is the use of measured time delays to infer $H_0$: this method requires building detailed mass models for gravitational lenses that host multiple images of background sources. These models, along with measured time delays, enable system-by-system constraints, which can be combined to get a better measurement of $H_0$, and even other cosmological parameters, from a catalog of such systems (see \cite{Birrer:2022chj} for a recent review). The dependence of lensing time delay on cosmological parameters is degenerate with the lens parameters and the source location, which are usually {difficult to precisely constrain} in the absence of an EM counterpart. Hence previous work~\cite{Liao:2017ioi} in GW time-delay cosmography relied on the existence of an EM counterpart. This requires at least one of the compact objects in the binary to be a neutron star, and the mass ratio to be moderate, effectively restricting this method to low-mass binaries only. The horizon distance of 3G detectors to such low-mass binaries is modest ($z \lesssim 2$). Further, even with the best EM telescopes, the detectability of faint EM counterparts such as kilonovae is restricted to smaller redshifts ({$z \lesssim$ 0.5})~\cite{Scolnic_2017}. These limitations would make this essentially a probe of cosmology at low redshifts. 

In this work, we propose a {\em statistical} probe of cosmology that uses population-level properties of a catalog of lensed GW detections to constrain cosmology. This method is related to previous proposals to statistically infer cosmological parameters from distributions of image separations \cite{1993ApJ...419...12K, 1996ApJ...466..638K, Kochanek:2004ua, 2007ApJ...658L..71C} and time delays \cite{2007ApJ...660....1O, 2009ApJ...706...45C, 2020MNRAS.498.2871H, 2021A&A...656A.153S} in lensed quasars. In contrast to the lensing of light, the angular separation of lensed GW signals in the sky is expected to be unresolvable even with 3G detectors. However, we will have precise timing information. Even with existing GW detectors, the arrival times of GW signals can be measured with a precision of $\sim10^{-4}$~s, which cannot be achieved using quasar light curves. Moreover, GWs are unaffected by issues such as extinction (which is a potential source of systematic error for quasar cosmography~\citep{Maoz:2005xt}) and in general have a much simpler and well-modeled selection function.

{We aim to} to look for the imprints of cosmological parameters on the {number} of lensed signals, as well as the distribution of their time delays, without relying on the accurate knowledge of the source location of the individual signals and the properties of the corresponding lenses. Indeed, the number of lensed events as well as the time delay distribution will also depend on the source properties (e.g., mass and redshift distribution of BBHs~\cite{haris2018,Li:2018prc,Mukherjee:2021qam}) as well as the lens properties (e.g., the mass function of the dark matter halos~\cite{2020MNRAS.495.3727R} and the lens model~\cite{2022arXiv220511499J,More:2021kpb}). If the distributions of the source and lens properties are known from other observations or theoretical models (e.g., from the observation of unlensed GW signals and dark matter simulations), then the cosmological parameters can be inferred from the observed number of lensed events and their time delay distribution. Recently \cite{Xu:2021bfn} presented a complementary approach where they sought to constrain the distribution of the sources and lenses from the GW lensing rate and time delay distribution.

We provide prospective constraints on $H_0, \Omega_{m}$ (and by extension, {$\Omega_{\Lambda}$, assuming a flat $\Lambda$CDM cosmology)} acquired from the expected detections of strongly lensed BBH detections in the 3G era. For simplicity, we set the rest of the parameters in the $\Lambda$CDM model to be the values given by the recent CMB measurements~\cite{Planck18}. We adopt a Bayesian framework which effectively compares the detected number of lensed events and the time delay distribution of lensed events, to a set of model distributions corresponding to different values of $H_0$ and $\Omega_{m}$. We assume that the distributions of the source properties and lens properties are known from independent means (e.g. from the population of unlensed GW signals, EM probes, or theoretical models). We find that we are able to constrain $H_0$ ($\Omega_m$) with a $\sim 1.6\%$ ($1.9\%$) precision ($68\%$ credible interval), assuming an observation period of 10 yrs and a BBH merger rate of $5 \times 10^5~\mathrm{yr}^{-1}$.

\section{Strong lensing probability and time delay}

GW signals that we consider have wavelengths much smaller than the characteristic length scales of the lenses (galaxies and clusters), which are, in turn, much smaller than the cosmological distances. Hence we invoke both the geometric optics approximation as well as the thin lens approximation~\cite{dodelson2017}. Further, we assume that the {lenses} can be approximated as {singular isothermal spheres (SIS)}, which is a reasonable first approximation to {galaxies} situated at the centre of spherical dark matter {halos}. Note that the method that we propose does not rely on this simple lens model --- {detailed} analyses will need to use more realistic lens models\footnote{See, e.g., \cite{haris2018,More:2021kpb} for time delay distributions for some other lens models. {Note also that our method is statistical in nature, and hence, we need a good description of lenses on average, rather than precise mass models of individual lenses.}}.

In the SIS approximation, if the source lies within the Einstein radius $r_E$ of the lens, it will produce two distinct images \cite{oguri2002} (equivalently, two copies of the GW signals that are potentially observable). For a source at redshift $\zs$, the probability of this happening is given by $P_\ell(z_s|\vec{\Omega}) = 1 - \exp[-\tau(z_s,\vec{\Omega})]$. Here, $\tau(z_s,\vec{\Omega})$ is the strong lensing optical depth --- a measure of the total number lenses that lie within a radius $r_E$ from the line connecting the source and the Earth. 
\begin{equation}\label{tau-defn}
\tau(z_s, \vOmega) =\int_{0}^{z_s} \frac{d\tau}{dz_{\ell}} (z_s, \vOmega) ~ d\zl.
\end{equation}
The differential optical depth ${d\tau}/{dz_{\ell}}$ has a dependence on cosmological parameters $\vOmega$. Hence the fraction of lensed GW events will {depend on} $\vOmega$.  

The time delay between the two images is given by (see, e.g. \cite{oguri2002}):
\begin{equation}\label{delta-t-sis}
\Delta t (\zl, \sigma, \zs, y, \vec{\Omega})  =  \frac{D_{\Delta t}}{c} ~{32\pi^2}\left(\frac{\sigma}{c}\right)^4 ~ y ~ \left(\frac{\Dls}{\Ds}\right)^2. 
\end{equation}
Above, $D_{\Delta t} \equiv (1+\zl) {\Ds \Dl} / {\Dls}$, is called the time delay distance, where $D_{\ell}$, $D_s$, $D_{\ell s}$ are angular diameter distances separating the earth and the lens, the earth and the source, and the lens  and the source, respectively. The lens and source redshifts are denoted by $\zl$ and $\zs$, while $\sigma$ denotes the velocity dispersion of the lens,  and $y$ the source position in the lens plane (in units of $r_E$). The time delay depends, in addition, on the cosmological parameters $\vec{\Omega}$.  This is because the relation between angular diameter distance and redshift depends on cosmological parameters. 

If the properties of the lens ($\sigma$ and $\zl$ in the SIS approximation) and the source ($z_s$ and $y$) are known (either through an EM counterpart of the merger or by localizing the host galaxy\cite{Hannuksela_2020}), then Eq.~\eqref{delta-t-sis} can be solved to get the cosmological parameters directly~\cite{Liao:2017ioi}. However, this is likely to be possible only for a small number of detected lensed events and limit ourselves to low redshifts ($z \lesssim 0.5$). Hence, we resort to a statistical approach here, where we exploit the dependence of the fraction of lensed events, as well as the time delay distribution of the images, on cosmological parameters. 

\section{Bayesian inference of cosmological parameters from lensed events}

\begin{figure*}[tbh]
\includegraphics[height=2.4in]{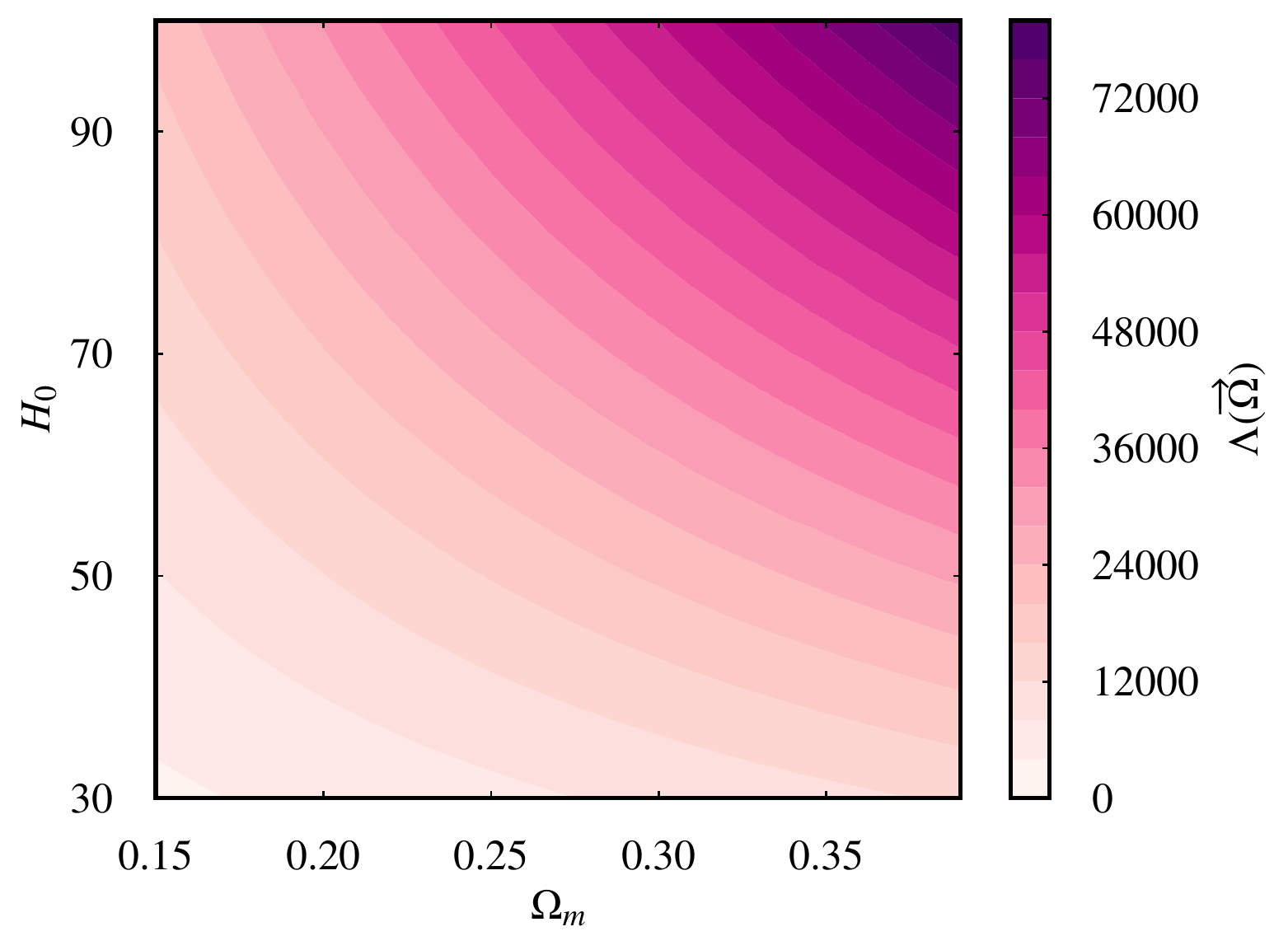}
\includegraphics[height=2.4in]{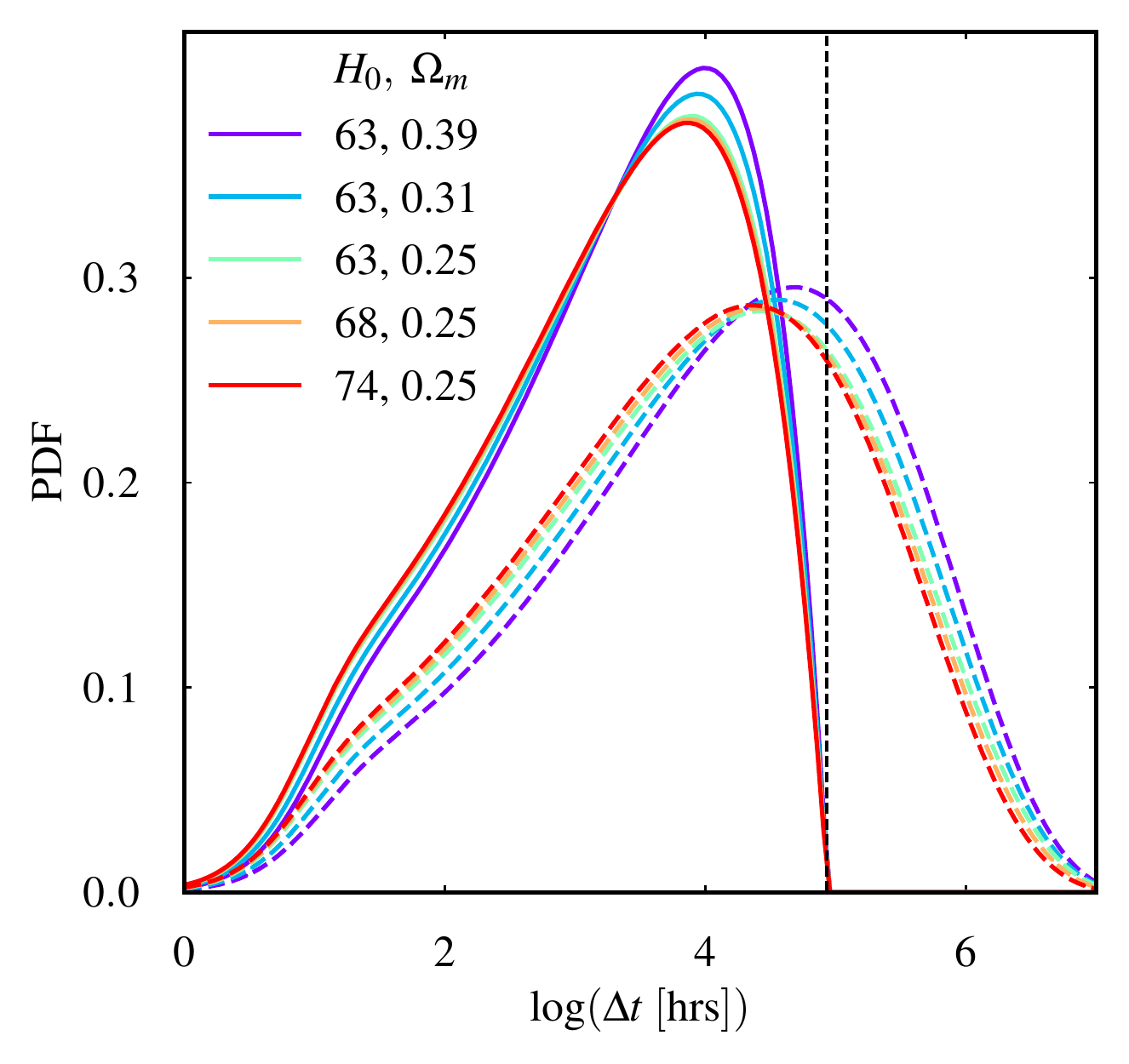}
\caption{{\it Left:} Expected number of lens pairs for different values of $\vec{\Omega}$ in flat $\Lambda$CDM model, assuming a merger rate $R = 5\times10^5\;\mathrm{yr}^{-1}$ and observation time period $T_{\mathrm{obs}}=10\;\mathrm{yrs}$.  {\it Right:} The strong-lensing time-delay distributions for different values of cosmological parameters: Increasing $H_0$ ($\Omega_m$) shifts the peak of the distribution towards smaller (larger) time-delay values. Dashed lines show the actual distributions $p(\Delta t~|~\vec{\Omega})$ while the solid lines show the distribution of time delays observable in a period of 10 yrs.}
\label{fig:delta-t}
\end{figure*}

We assume that a set of $N$ lensed BBH mergers have been confidently detected within an observation period $T_{\mathrm{obs}}$, each merger producing two lensed copies of the GW signal. We also assume that the time delays between each of these the lensed pairs have been measured accurately and precisely, which we denote as $\lbrace \Delta t_i \rbrace, ~ i = 1, 2, \ldots, N$. Given $N$ and $\lbrace \Delta t_i \rbrace$, we wish to construct a posterior on cosmological parameters $\vec{\Omega} \equiv (H_0, \Omega_m)$ assuming a flat $\Lambda$CDM cosmology.  From Bayes' theorem:
\begin{equation}
p\left(\vec{\Omega}~|~N,\lbrace \Delta t_i \rbrace \right) = \frac{p\left(\vec{\Omega} \right)p\left(N,\lbrace \Delta t_i \rbrace~|~\vec{\Omega},T_{\mathrm{obs}} \right)}{Z}
\label{eq:posterior}
\end{equation}
where $p\left(\vec{\Omega} \right)$ is the prior on $\vec{\Omega}$ and $p\left(N,\lbrace \Delta t_i \rbrace~|~\vec{\Omega},T_{\mathrm{obs}} \right)$ is the likelihood of observing $N$ lensed events with time delays $\lbrace \Delta t_i \rbrace$ given the cosmological parameters $\vec{\Omega}$.  $Z$ is a normalization constant, which is the evidence of the assumed model.

Since $N$ and $\lbrace \Delta t_i \rbrace$ are uncorrelated, the likelihood can be written as a product of likelihoods:
\begin{equation}
p\left(N,\lbrace \Delta t_i \rbrace~|~\vec{\Omega},T_{\mathrm{obs}} \right) = p\left(N ~|~\vec{\Omega},T_{\mathrm{obs}}\right) p\left(\lbrace \Delta t_i \rbrace ~|~\vec{\Omega},T_{\mathrm{obs}} \right).
\end{equation}
Above, the likelihood of observing $N$ lensed BBH mergers can be described by a Poisson distribution
\begin{equation}
\label{eq:Nlikelihood}
	p\left(N ~|~\vec{\Omega},T_{\mathrm{obs}}\right) = \dfrac{\Lambda(\vec{\Omega},T_\mathrm{obs})^N\;e^{-\Lambda(\vec{\Omega},T_{\mathrm{obs}})}}{N!},
\end{equation}
where $\Lambda(\vec{\Omega},T_{\mathrm{obs}})$ is the expected total number of lensed events observed within the observation period as predicted by the cosmological model with parameters $\vec{\Omega}$. Assuming that different BBH mergers are independent events, the likelihood for observing the set of time delays $\lbrace \Delta t_i \rbrace$ can be written as
\begin{equation}
	p\left(\lbrace \Delta t_i\rbrace~|~\vec{\Omega},T_{\mathrm{obs}}\right) = 
	\prod_{i=1}^{N} p\left(\Delta t_i~|~\vec{\Omega},T_{\mathrm{obs}}\right). 
\end{equation}
$p(\Delta t_i~|~\vec{\Omega},T_{\mathrm{obs}})$, can be thought of as ``model'' time-delay distribution $p(\Delta t~|~\vec{\Omega},T_{\mathrm{obs}})$  evaluated at the measured $\Delta t_i$, whose shape is governed by the cosmological parameters $\vec{\Omega}$. $p(\Delta t~|~\vec{\Omega},T_{\mathrm{obs}})$ is obtained from the expected time delay distribution $p(\Delta t~|~\vec{\Omega})$, after applying the condition that we can not observe the time delays which are greater than the observation time $T_{\mathrm{obs}}$:
\begin{equation}\label{selection-function}
p\left(\Delta t~|~\vec{\Omega},T_{\mathrm{obs}}\right) \propto p\left(\Delta t~|~\vec{\Omega}\right) {\left(T_{\mathrm{obs}}-\Delta t\right)} \, \Theta( T_\mathrm{obs} - \Delta t),\\
\end{equation}
where $\Theta$ denotes the Heaviside function.  
We evaluate the posterior $p(\vec{\Omega}~|~N,\lbrace\Delta t_i\rbrace)$ on a finely meshed grid spanning the space of cosmological parameters $\vec{\Omega}$. The likelihood $p(N,\lbrace\Delta t_i\rbrace~|~ \vec{\Omega},T_{\mathrm{obs}})$ requires the calculation of the expected total number of lensed events $\Lambda(\vec{\Omega},T_{\mathrm{obs}})$ and the expected time delay distribution $p(\Delta t~|~ \vec{\Omega})$ for different values of $\vec{\Omega}$. Indeed these quantities depend on the distribution of the source
and lens properties, such as the redshift distribution of BBH mergers and the halo mass function. In this work, we assume that these properties are known from other observations.

We compute the expected number of lensed binaries using the following integral
\begin{equation}\label{eq:Lambda}
	\begin{split}
	\Lambda(\vec{\Omega},T_{\mathrm{obs}}) = R & \int_0^{z_s^\mathrm{max}} {p_b}(z_s | \vec{\Omega})  \, P_\ell(~z_s|\vec{\Omega}) \,  d z_s\\
	&\times \int_{\Delta t=0}^{T_{\mathrm{obs}}} p(\Delta t|\vec{\Omega}) \, (T_{\mathrm{obs}}-\Delta t) \, d\Delta t,
	\end{split}
\end{equation}
where $R$ is the BBH merger rate, $p_{b}(z_s | \vec{\Omega})$ is the redshift distribution of merging binaries and  $P_\ell(z_s | \vec{\Omega})$ is the strong lensing probability for the source redshift $z_s$.  Here we assume that the GW detectors are able to detect all the merging binaries out to $z_{\mathrm{max}}$. For 3G detectors, this is a good assumption for the $z_\mathrm{max}$ values that we use~\footnote{The $z_{\mathrm{max}}$ predicted by a source population model (e.g., \cite{dominik2013}) assumes the standard cosmology   $\vOmega_\mathrm{true}$. When we consider other values of $\vOmega$, we rescale  $z_\mathrm{max}$ appropriately.}.

Similarly, we compute the expected time delay distribution $p (\Delta t~|~\vec{\Omega} )$ for different values of the cosmological parameters $\vec{\Omega}$ by marginalising the distribution of time delay over all other parameters $\vec{\lambda} \equiv \{y,\sigma,z_{\ell},z_s\}$ on which the time delay depends [see Eq.(\ref{delta-t-sis})]:
\begin{equation}\label{delta-t-integration}
	p\left(\Delta t ~|~ \vec{\Omega}\right) = \int p\left(\Delta t ~|~ \vec{\lambda},\vec{\Omega}\right) p (\vec{\lambda}~|~\vec{\Omega}) \, d\vec{\lambda},
\end{equation}
where $p(\vec{\lambda}~|~\vec{\Omega})$ denotes the expected distribution of the impact factor $y$, lens velocity dispersion $\sigma$, lens redshift $z_{\ell}$ and source redshift $z_s$, given the cosmological parameters $\vec{\Omega}$.

We assume that redshift distribution $p_b(\zs~|~\vec{\Omega})$ will be known with adequate precision from the observation of the larger number of unlensed events, which will dominate the data. For illustration, we take the model described in \cite{dominik2013} as the true model of $p_b(\zs)$. The lensing optical depth $\tau(z_s,\vec{\Omega})$ depends on the source redshift $\zs$, the assumed cosmology $\vec{\Omega}$, and a model of the lens distribution [see Eq.\eqref{tau-defn}]. We model the lens distribution using the halo mass function, which gives the distribution $p(\sigma, \zl)$ of $\sigma$ and $\zl$. 
We  consider the halo mass function model described in \cite{behroozi2013}, but use an additional model \cite{jenkins2001} to check the bias introduced by using a wrong model in the parameter inference (see Supplemental Material).
 Finally, the distribution of  impact parameter $y$, $p_y(y) \propto y$, with $y \in [0, 1]$. This corresponds to a uniform distribution of lensed sources in the lens plane within the Einstein radius. 

Figure~\ref{fig:delta-t} illustrates the imprint of cosmology on the number of lensed events observable for a period of ten years as well as the distribution of time delays. The number of lensed events increase with increasing $H_0$ and $\Omega_m$. The peak of the distribution shifts towards smaller time-delay values with increasing $H_0$, and towards larger values with increasing $\Omega_m$. Even though the impact of varying cosmology on the time delay distribution appears small by eye, the Bayesian approach delineated in this section is able to adequately capture these imprints to provide $\mathcal{O}(1\%)$ constraints.

\section{Expected constraints on cosmological parameters}\label{sec:results}

\begin{figure}[tbh]
\includegraphics[width=1.64in]{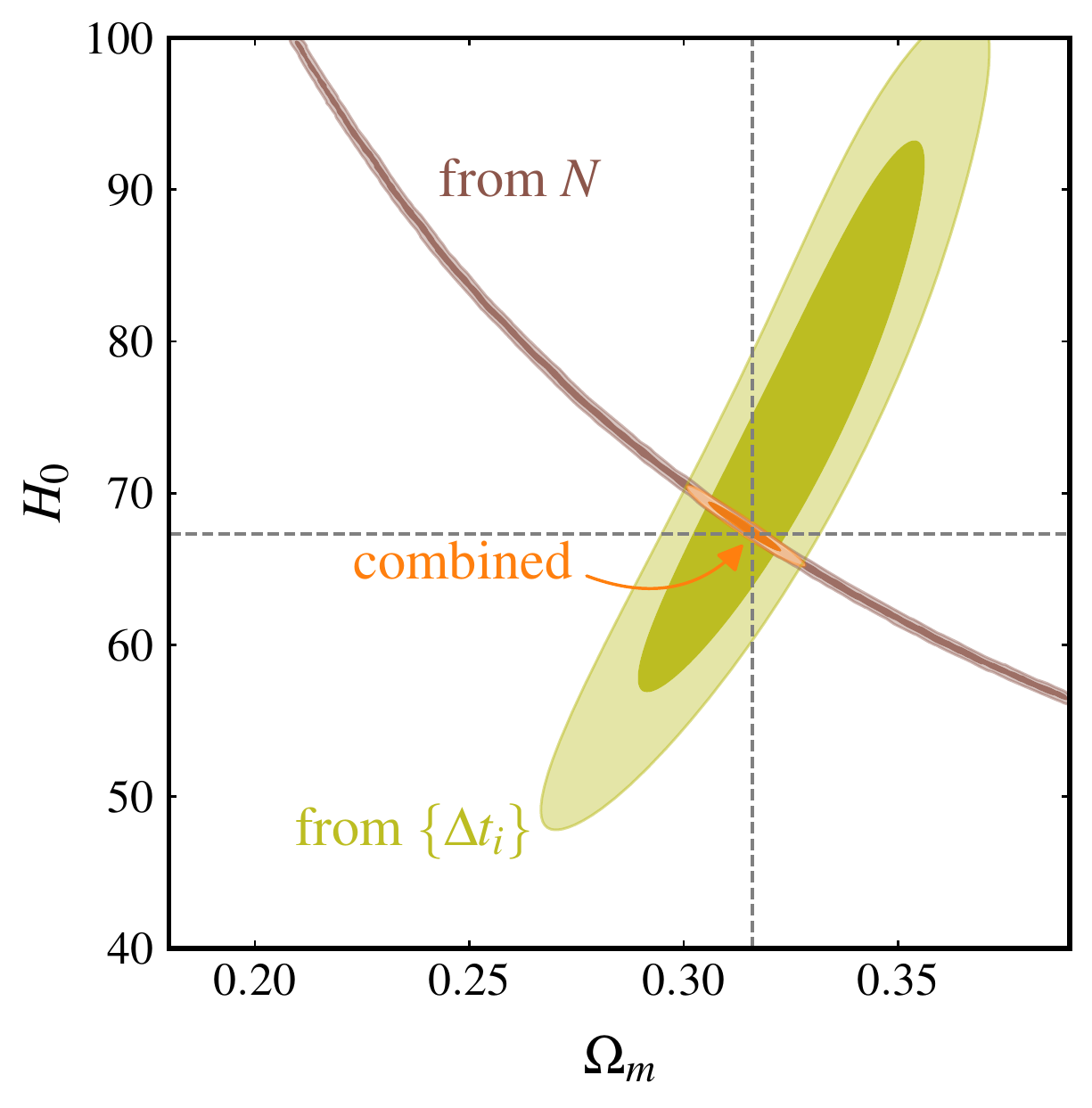}
\includegraphics[width=1.64in]{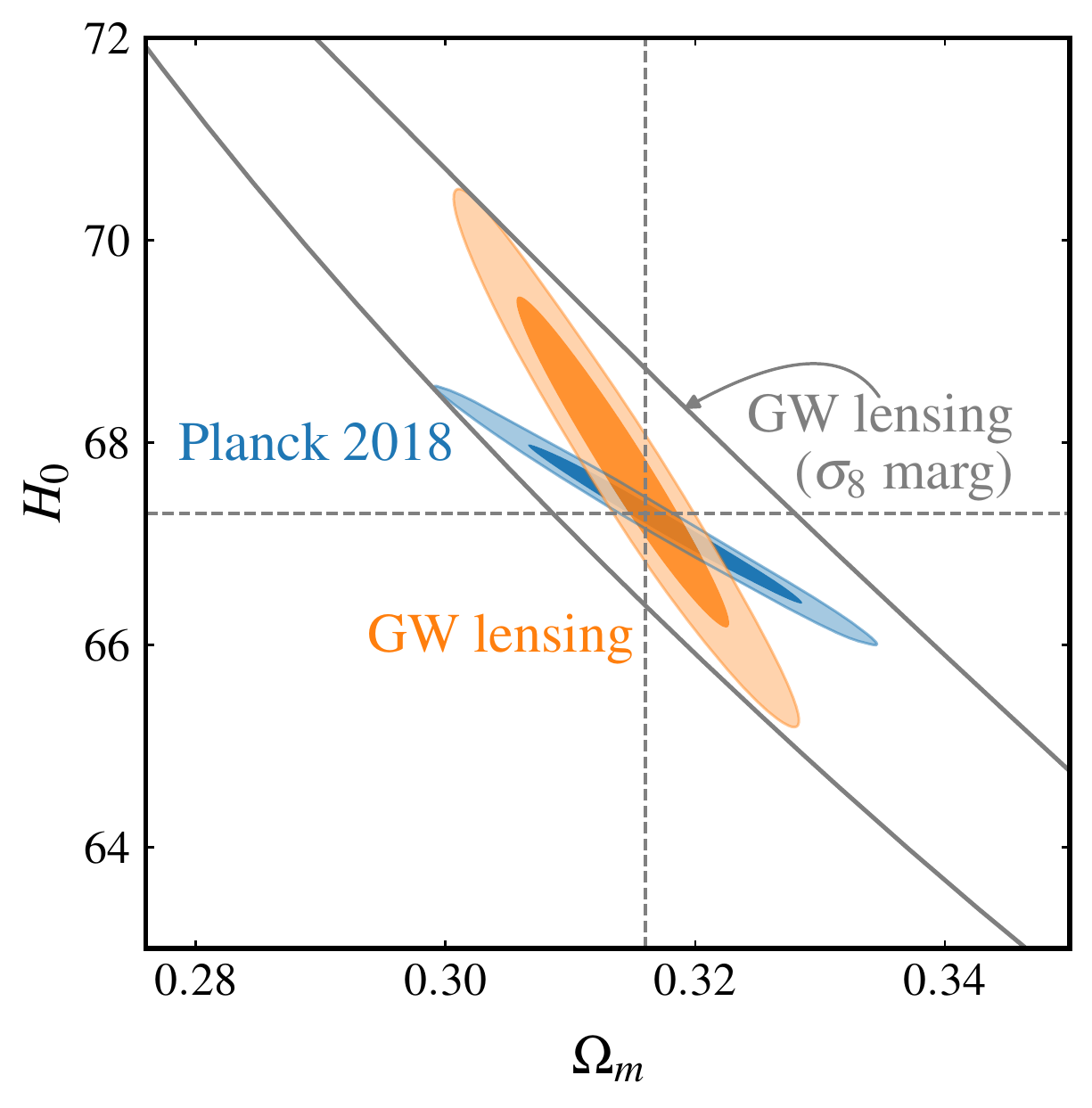}
\caption{\emph{Left panel:} Expected posterior distributions (68\% and 95\% credible regions) of $H_0$  and $\Omega_m$ computed from the time delay distribution and the number of lensed events separately, along with the combined posterior (shown in orange). We assume a BBH merger rate $R=5\times10^5\;\mathrm{yr^{-1}}$ and total observation time period $T_{\mathrm{obs}}=10\;\mathrm{yrs}$. The ``true'' cosmology (dashed cross-hairs) is recovered within the $68\%$ credible interval (orange shade), with $H_0 = 67.8 \pm 1.1~\mathrm{km ~s^{-1} Mpc^{-1}}$ and $\Omega_m = 0.3142 \pm 0.0056$. \emph{Right panel:} A comparison of the the combined posterior obtained from GW lensing with the same obtained from CMB observations by Planck. While the orange contours assume that the $\sigma_8$ parameter is well-measured from other observations, the grey contour corresponds to $95\%$ credible region of the posterior marginalized over $\sigma_8$ parameter. }
\label{fig:lensing-planck}
\end{figure}

To assess the ability of our method to constrain cosmological parameters, we choose a ``true'' cosmology $\vec{\Omega}_{\mathrm{true}} = \{H_0 = 67.3,\; \Omega_m = 0.316\}$.  We further assume that the ``true'' halo mass model is described by \cite{behroozi2013}, as implemented in the \textsc{hmfcalc} package \cite{hmfcalc}, and the ``true'' source distribution is given by~\cite{dominik2013}. We assume a total observing period $T_{\mathrm{obs}}=10\;\mathrm{yrs}$ and a BBH merger rate $R=5\times10^5\;\mathrm{yr}^{-1}$. We neglect the selection effects in the detection as 3G detectors are expected to detect all the BBHs out to large distances ($d_L \sim 1000$~Gpc). We compute the expected number $\Lambda$ of lensed events making use of Eq.\eqref{eq:Lambda}. To simulate one observational scenario where $N$ events are detected, we draw one sample from a Poisson distribution with mean $\Lambda$. Further, we draw samples $\lbrace \Delta t_i \rbrace_{i=1}^N$ from $p(\Delta t~|~\vec{\Omega}_{\mathrm{true}},T_{\mathrm{obs}})$ [see Eq.\eqref{selection-function}].

Using $N$ and $\lbrace \Delta t_i \rbrace_{i=1}^N$, we evaluate the posterior described in Eq.~\eqref{eq:posterior} for different values of $\vec{\Omega}$. We assume uniform priors on $H_0$ and $\Omega_m$, so that the final posterior is given by the product of the likelihoods $p(N~|~\vec{\Omega},T_{\mathrm{obs}})$ and $p({\Delta t_i}~|~\vec{\Omega},T_{\mathrm{obs}})$. Figure~\ref{fig:lensing-planck} shows these two likelihoods as well as the posterior on $H_0$ and $\Omega_m$ obtained from combining these two likelihoods. We find that the posteriors are centred around the true values of cosmological parameters. Further, the constraints on $\vec{\Omega}$ are found to be  $H_0 = 67.8 \pm 1.1$ and $\Omega_m = 0.314 \pm 0.006$ (68\% credible intervals of marginalised posteriors). These constraints are comparable to those derived from {the} CMB \cite{Planck18}~\footnote{Note that we have set all other parameters of the $\Lambda$CDM model to the best fits values provided by \cite{Planck18}. In order to make a fair comparison, we do the same for the Planck posteriors {as well}. However, the uncertainty in some of the other parameters, in particular $\sigma_8$, will have an imprint on the precision with which $H_0$ and $\Omega_m$ could be constrained. Therefore, in addition, we show the posteriors that are marginalized over $\sigma_8$ as well in Fig. \ref{fig:lensing-planck}. The marginalized constraints are significantly worse, so we need a complimentary probe to achieve better constraining power.}. Additionally, they probe a very different redshift regime ($z\sim10$  as compared to $z \sim 1000$ probed by the CMB) and have different systematics. 

\begin{figure}[tbh]
\includegraphics[scale=0.432]{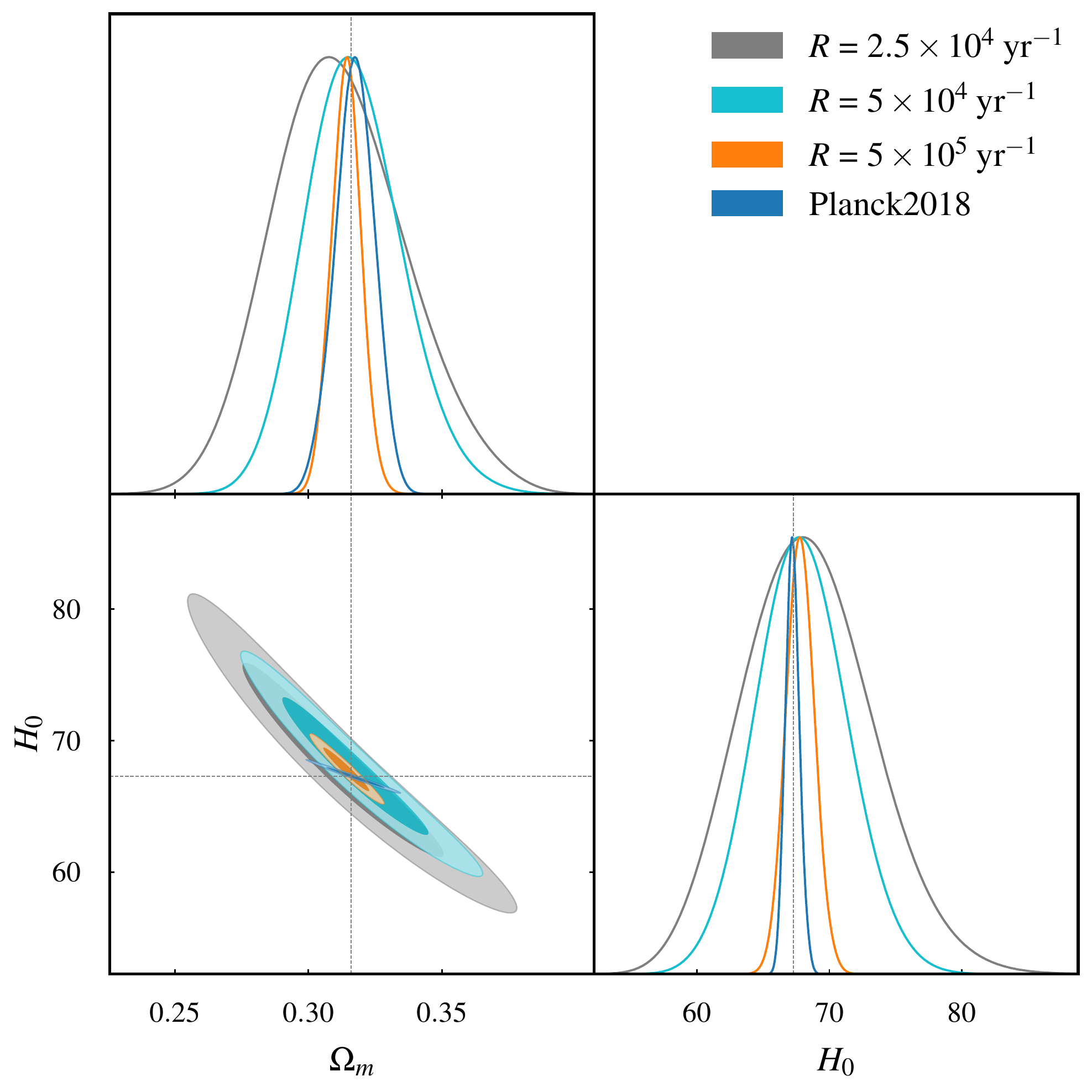}
\caption{Expected posterior distributions of $H_0$  and $\Omega_m$ from a 10-year observation period, assuming different values for the merger rate $R$ (shown in the legend). A lower merger rate (producing a smaller number of lensed events) will result in less precise estimates of the cosmological parameters.}
\label{fig:lensing-samples}
\end{figure}

While we have assumed a BBH merger rate of $R = 5\times10^5\;\mathrm{yr}^{-1}$, the true merger rate is uncertain as of now. Hence we repeat these calculations assuming a more moderate merger rate of $R = 5\times10^4\;\mathrm{yr}^{-1}$ and a pessimistic rate of $R = 2.5\times10^4\;\mathrm{yr}^{-1}$. This will, in turn reduce the observed number of lensed events over the observational period of $T_{\mathrm{obs}} = 10\;\mathrm{yrs}$. The expected posteriors on cosmological parameters assuming the three different merger rates is shown in Fig.\ref{fig:lensing-samples}. As expected, a smaller merger rate (resulting in a smaller number of lensed events) increases the width of the posteriors, although the true cosmology continues to be recovered within the $68\%$ credible interval.

\begin{figure}[tbh]
\includegraphics[scale=0.4]{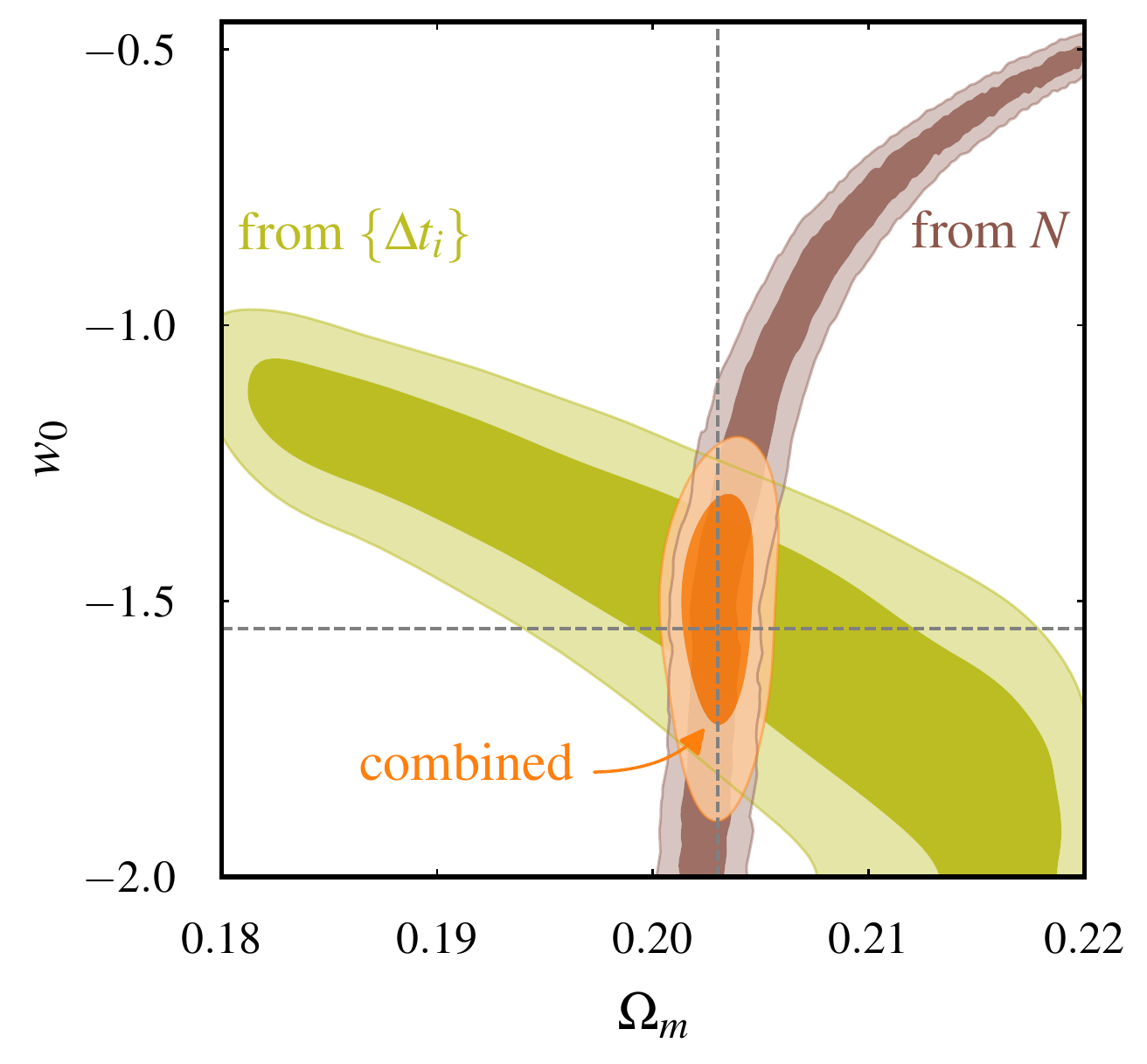}
\caption{Expected posterior distributions (68\% and 95\% credible regions) of $\Omega_m$ and $w_0$ of the $w$CDM model computed from the time delay distribution and the number of lensed events (jointly). We assume a BBH merger rate $R=5\times10^5\;\mathrm{yr^{-1}}$ and total observation time period $T_{\mathrm{obs}}=10\;\mathrm{yrs}$. The ``true'' cosmology (dashed cross-hairs) is recovered within the $68\%$ credible interval (orange shade), with $w_0 = -1.52^{+0.16}_{-0.12}$ and $\Omega_m = 0.203 \pm 0.001$.}
\label{fig:lensing-planck-wcdm}
\end{figure}

We also illustrate the ability of this method to constrain parameters of some more general cosmological models. In particular, we consider $w\mathrm{CDM}$ model~\cite{Planck18} with two  parameters $\vec{\Omega}= \{\Omega_m,\;w_0 \}$. In this part, we fix the Hubble constant $H_0$ to its ``true'' value, mimicking a situation where it will be well measured from low-redshift observations. As done earlier, we compute the expected number of lensed events and model time delay distributions using Eqs.~\eqref{eq:Lambda} and \eqref{selection-function}. We choose a ``true" cosmology  $\vec{\Omega}_{\mathrm{true}}=\{\Omega_m = 0.203,\;w_0 = -1.55\}$. We assume a halo mass model described in \cite{behroozi2013} and the source distribution given in \cite{dominik2013}. Now considering a BBH merger rate $R = 5\times10^5\;\mathrm{yr}^{-1}$  and total observing time period $T_{\mathrm{obs}}=10\;\mathrm{yrs}$, we draw one value of $N$ and  one set of$\lbrace\Delta t_i\rbrace_{i=1}^N$ from $p(N ~|~\vec{\Omega}_\mathrm{true},T_\mathrm{obs})$ and $p(\Delta t~|~\vec{\Omega}_\mathrm{true}, T_\mathrm{obs})$. From these simulated observation data we evaluate the posteriors on these $\Omega_m$ and $w_0$  (see Fig.~\ref{fig:lensing-planck-wcdm}). {The expected} constraints from GW lensing ($w_0 = -1.52^{+0.16}_{-0.12}$ and $\Omega_m = 0.203 \pm 0.001$) {compare favorably to those} obtained from Planck ($w_0 = -1.55^{+0.18}_{-0.33}$ and $\Omega_m = 0.203_{-0.058}^{+0.018}$){, albeit with the caveat that we are exploring only a subset of parameters}.

\section{Outlook}\label{sec:conclusion}

We expect the lensing cosmography to provide a complementary measurement of cosmological parameters that are {comparable to those derived from} other cosmological probes, and at the same time {using data from} an intermediate regime in redshift ($z \sim 10$) that is rarely explored by other probes. The apparent tension that exists between the current low and high redshift measurements underlines the need of additional measurements -- especially the ones probing an intermediate redshift regime. 

The major sources of systematic errors that affect quasar cosmography~\cite{Maoz:2005xt} are unlikely to affect GW lensing cosmography: GWs are unaffected by extinction, {and} selection effects in GW searches are better modelled thanks to their intrinsic simplicity. Nevertheless, there are several challenges to overcome before this method {can} be used to provide reliable measurements of cosmological parameters.  The number of lensed events as well as the distribution of their time delays depend on the properties of the astrophysical sources and lenses. Some of the relevant parameters, such as the redshift distribution of BBH mergers, {can} be inferred from the unlensed GW signals and the stochastic GW background. For other parameters, such as the distribution of the properties of the lens parameters, will need to rely on theoretical models, such as large-scale cosmological simulations. 

There are other potential sources of systematics: the detected sample of lensed events will also contain some unlensed events that are misidentified as lensed, thus contaminating our measurement. This contamination fraction will increase with increasing number of detections~\cite{Caliskan:2022wbh}. However, in third generation detectors, we will have better ability distinguish between lensed and unlensed events (e.g., due to the improved precision in estimating source parameters, by measuring additional signal properties such as spin angles, higher harmonics and Morse phase, and by using better informed priors on source parameters, lensing time delay and magnification). Thus, we expect that it will be possible to keep the contamination fraction sufficiently small \cite{Janquart:2022wxc,Janquart:2023osz,More:2021kpb}(see Supplemental Material). In addition, even this contamination can be modelled as the unlensed events will have a different, known time delay distribution (see, e.g., Fig .2 of~\cite{haris2018}). 

In this paper, we have assumed a simple lens model, considered only two cosmological parameters as free and have employed an approximate treatment of the calculation of the lensing optical depth. While this paper was meant to illustrate this idea and to demonstrate the potential power of this approach, all these approximate treatments need to be revisited and refined. Additionally, we have focussed on BBH signals, although BNS and NSBH mergers could also be used for these measurements. Hence the expected constraints that we report here are only indicative of the potential of this method. While challenges are great, potential payoffs makes this a worthwhile pursuit.  We will explore this research program in {future work}.

\section{Acknowledgements} \label{sec:ack}

We are grateful to Surhud More and the anonymous referees for their careful review of the manuscript and useful comments. We are grateful to Steven Murray for his valuable assistance with the HMFcalc package. We also thank the members of the ICTS Astrophysics \& Relativity group, in particular Aditya Vijaykumar, for useful discussions. Our research was supported by the Department of Atomic Energy, Government of India, under Project No. RTI4001. SJK’s work was supported by a grant from the Simons Foundation (677895,  R.G.).  PA’s research was supported by the Canadian Institute for Advanced Research through the CIFAR Azrieli Global Scholars program.   {TV acknowledges support by the National Science Foundation under Grant No. 2012086.} All computations were performed with the aid of the Alice computing cluster at the International Centre for Theoretical Sciences, Tata Institute of Fundamental Research.

\bibliography{references}

\clearpage
\onecolumngrid 
\begin{center}
\textbf{\large Supplemental Material}
\end{center}
\twocolumngrid 

\setcounter{equation}{0}
\setcounter{figure}{0}
\setcounter{table}{0}
\setcounter{page}{1}

\section{Constructing template time-delay distributions}\label{templates}

We call $p(\Delta t~|~\vec{\Omega})$  the intrinsic time delay distribution and $p(\Delta t ~|~\vec{\Omega},T_{\mathrm{obs}})$  the model/template time delay distribution (see Eqs.~(0.9)
and (0.7)
, respectively in the main text). The model distribution $p(\Delta t ~|~\vec{\Omega},T_{\mathrm{obs}})$ is obtained from the intrinsic distribution $p(\Delta t~|~\vec{\Omega})$ after applying the fact that we can not observe time delays greater than the total observation time period $T_{\mathrm{obs}}$. Here we briefly summarize the key steps in obtaining the intrinsic time delay distribution.

In the singular isothermal sphere (SIS) approximation of the lens, the image time-delay $\Delta t$ depends on the redshift of the source $\zs$ and the lens $\zl$, the velocity dispersion of the lens $\sigma$ and the impact parameter $y$, apart from the cosmological parameters $\vec\Omega$ (cf. Eq. 0.2, in the main text
).  We are interested in the time delay distribution $p(\Delta t~|~\vec{\Omega})$ that is marginalised over $\zl, \sigma, \zs$ and $y$: 
\begin{equation}\label{time-delay-dist-integration}
	\begin{split}
		p(\Delta t~|~\vOmega)  = \int p(\Delta t~|~\sigma, \zl, \zs, y, \vOmega) ~ p(\sigma, \zl, \zs ~ | ~ \Omega)  \\ 
		~ \times ~ p(y) ~ d\zs d\zl d\sigma dy, 
	\end{split}
\end{equation}
where $p(y) \propto y$ with $y \in [0, 1]$. Since the distribution of $y$ is independent of the other parameters, it is convenient to define a scaled time-delay $\bar{\Delta t} := \Delta t/y$, so that 
\begin{equation}\label{time-delay-dist-bar}
	\begin{split}
		p(\Delta t~|~\vOmega)  = \int p(y) ~ \bar{p}({\Delta t}/{y}~|~\vOmega ) ~ \frac{1}{y} ~ dy,
	\end{split}
\end{equation}
where $\bar{p}(\bar{\Delta t}~|~\vOmega)$ denotes the distribution of $\bar{\Delta t}$ marginalised over $\zl, \sigma$ and $\zs$
\begin{equation}
	\begin{split}
		\bar{p}(\bar{\Delta t}~|~\vOmega)  =  \int \delta\left( \bar{\Delta t} - \bar{\Delta t}(\sigma, \zl, \zs, \vOmega)\right) ~ p (\sigma, \zl, \zs ~ | ~ \vOmega) \\ 
		~ \times ~ d\zs d\zl d\sigma, 
	\end{split}
\end{equation}
where $\delta$ is the Dirac delta function. Above, $p(\zl, \sigma, \zs ~ | ~ \vOmega)$ can be further split as 
\begin{equation}
	p(\sigma, \zl, \zs ~ | ~ \vOmega) = p(\sigma, \zl, ~ | ~ \zs , \Omega)  ~ p(\zs ~ | ~ \vOmega), 
	\label{eq:P_sigma_zl_zs_given_Omega}
\end{equation}
where $p(\sigma, \zl, ~ | ~ \zs , \Omega)$ is estimated from the differential optical depth  
\begin{equation}
	p(\sigma, \zl, ~ | ~ \zs , \vOmega) \propto \frac{d\tau}{d\zl d\sigma} (\zs, \vOmega). 
	\label{eq:P_sigma_zl}
\end{equation}
The calculation of the differential optical depth is discussed in the next section. 

\begin{figure*}[tbh]
	\includegraphics[height=2.2in]{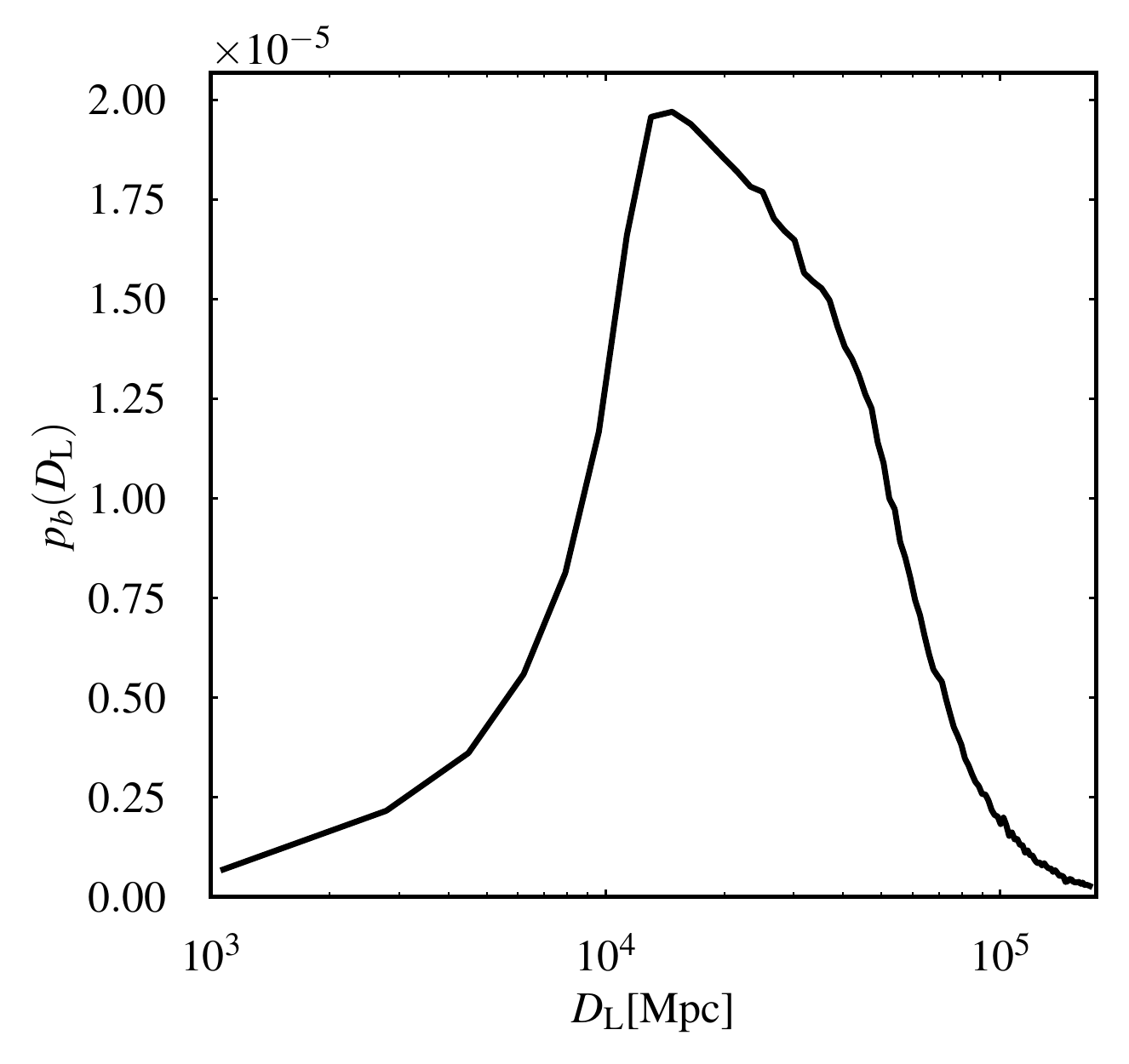}
	\includegraphics[height=2.1in]{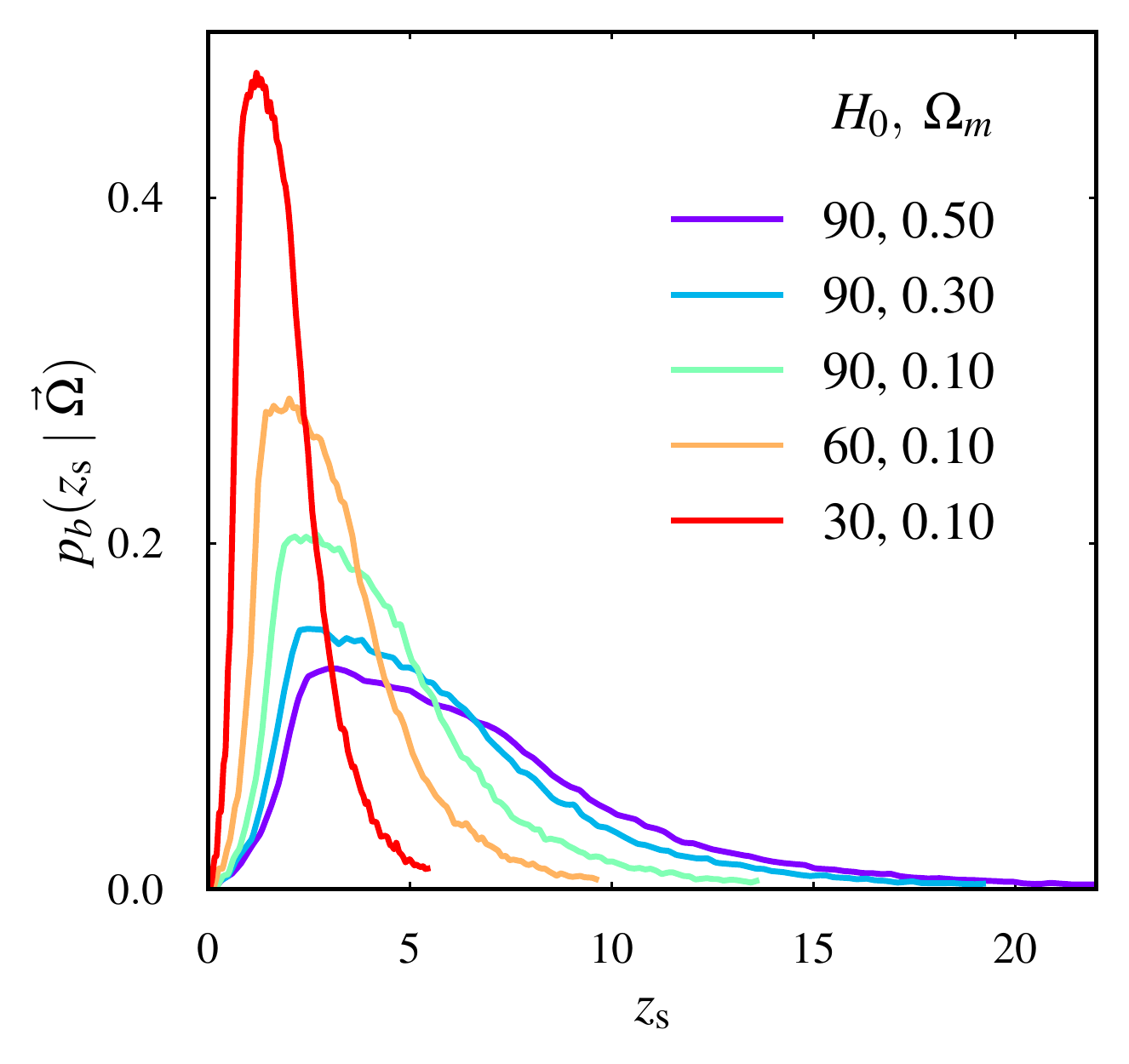}
	\includegraphics[height=2.1in]{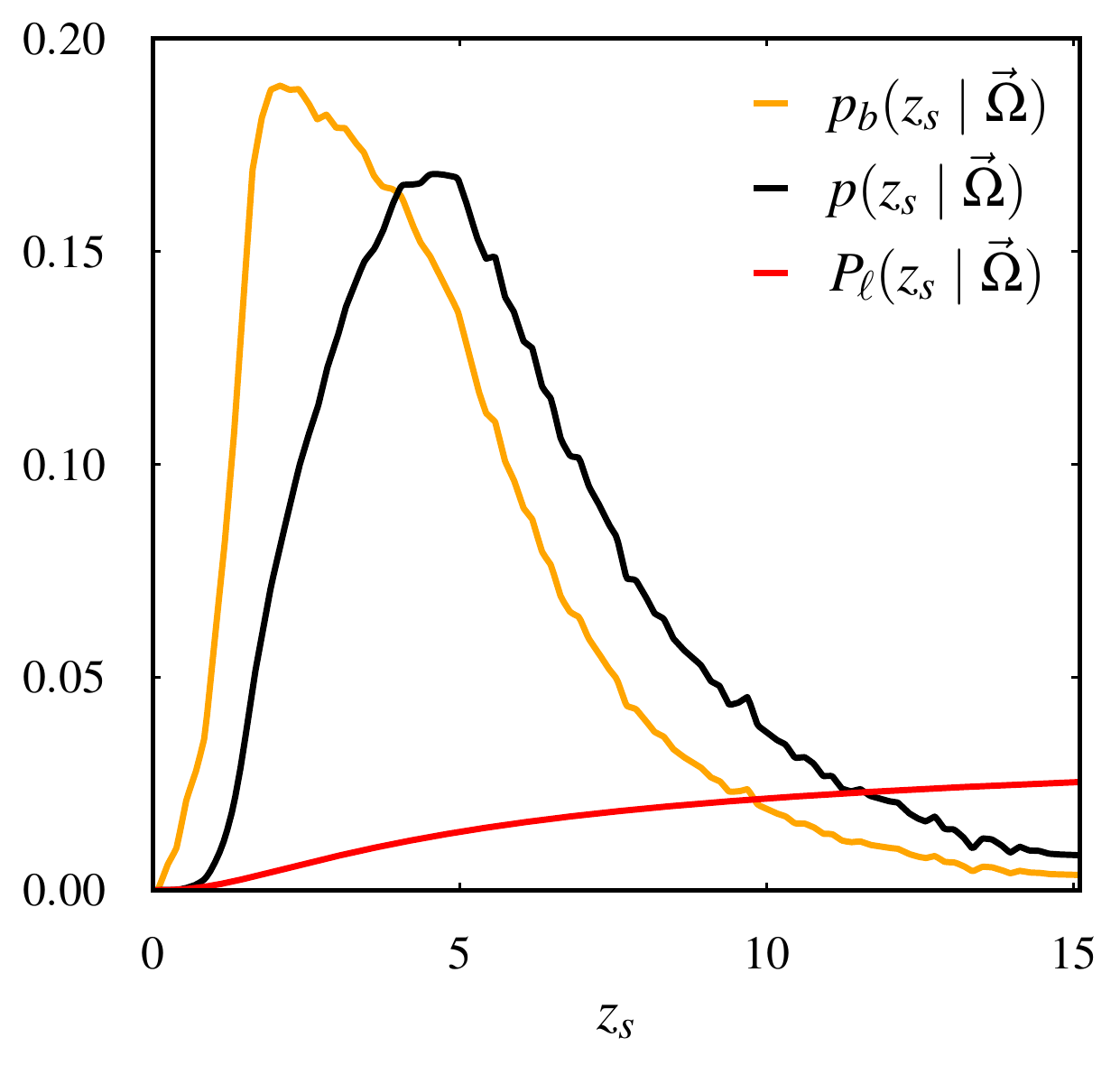}
	\caption{{\it Left Panel}: The redshift distribution of  BBH mergers from \cite{dominik2013} converted to a luminosity distance distribution, assuming standard cosmology \cite{Planck18}. {\it Middle Panel}: Source redshift distributions, converted from the aforementioned luminosity distance distribution, assuming different values of $\vec{\Omega}$. {\it Right Panel}: The source redshift distribution $p_b(z_s~|~\vec{\Omega})$ from \cite{dominik2013}, and the lensing probability $P_\ell(z_s~|~\vec{\Omega})$ assuming the halo mass-function described in \cite{behroozi2013}. The competing effects of a decreasing probability of sources and increasing probability of lensing at high redshifts  is reflected in the shifted peak of the distribution of lensed sources $p(z_s~|~\vec{\Omega})$ to higher redshifts. Here $\vec{\Omega}=\{H_0=67.3\;\mathrm{km}\;\mathrm{s}^{-1}\mathrm{Mpc}^{-1},\;\Omega_m = 0.316\}$.}
	\label{fig:dL-to-z}
\end{figure*}

\paragraph{Computing the redshift distribution of lensed BBHs}: In Eq.\ref{eq:P_sigma_zl_zs_given_Omega}, $p(\zs ~ | ~ \vOmega)$ is the redshift distribution of the lensed BBH systems, which is the product of the redshift distribution of merging BBHs $p_b(\zs ~ | ~ \vOmega)$ and the strong lensing probability $P_\ell(\zs ~ | ~ \vOmega)$ at that redshift: 
\begin{equation}
	p(\zs~ | ~ \vOmega) \propto p_b(\zs ~ | ~ \Omega) ~ P_\ell(\zs ~ | ~ \vOmega).
\end{equation}
The expectation is that the distribution of merging BBHs $p_b(D_L)$ as a function of the luminosity distance $D_L$ will be accurately measured from the large number of unlensed GW signals observed by 3G detectors. However, converting this into a redshift distribution $p_b(\zs ~ | ~ \vOmega)$ requires an assumption of a cosmology. For this work, we draw from a fiducial luminosity distance distribution obtained from \cite{dominik2013} (assuming the standard $\Lambda$CDM cosmology with parameters $\vOmega_{\mathrm{true}} = \{H_0 = 67.3,\; \Omega_m = 0.316\}$). This fiducial distribution $p_b(D_L)$ can be converted into redshift distributions $p_b(\zs ~ | ~ \vOmega)$ assuming different values for the cosmological parameters $\vOmega$ (see Fig.~\ref{fig:dL-to-z}). 

The lensing probability is given by $P_\ell(\zs ~ | ~ \vOmega) = 1-\exp[-\tau(\zs, \vOmega)]$, where $\tau(\zs, \vOmega)$  is the strong lensing optical depth, which can be computed by integrating the differential optical depth over the lens redshift and velocity dispersion 
\begin{equation}\label{tau-defined}
	\tau(z_s, \vOmega) = \int_{\sigma_\mathrm{min}}^{\sigma_{\mathrm{max}}}\int_{0}^{z_s} \frac{d\tau}{dz_{\ell}d\sigma} (z_s, \vOmega) ~ d\zl d\sigma.
\end{equation}
This prescription allows us to compute $p(\Delta t~|~\vec{\Omega})$ for different values of cosmological parameters $\vOmega$. 

\section{Calculation of the differential optical depth}\label{sec:appendix1}

The probability that the GWs from the source will encounter a lens --  and thus be strongly lensed -- on its trajectory towards the Earth is computed from the strong lensing optical depth $\tau$. The optical depth maybe thought of as the average number of lenses that lie within a proper distance $r_E(z_{\ell}, z_s)$, called Einstein radius, perpendicular to the line connecting the Earth and the source. 

In the SIS model of the lenses, the Einstein radius is given by:
\begin{equation}
	r_E(z_{\ell}, z_s) = 4\pi\left(\frac{\sigma}{c}\right)^2 \frac{D_{\Delta t}}{1+z_{\ell}} \left(\frac{D_{\ell s}}{D_s}\right)^2
\end{equation}
where $\sigma$ is the line-of-sight velocity dispersion of the particles that constitute the lens, $D_\mathrm{\Delta t} \equiv (1+z_{\ell})\frac{\Ds \Dl}{\Dls}$ is called the time delay distance   while $\Dl$ and $\Ds$ are the angular diameter distances to the lens and to the source (from the Earth) and $\Dls$ is the angular diameter distance between the lens and the source.  

The differential optical depth, at a lookback time $t_{\ell}$ corresponding to the location $\ell$, is a measure of the number of lenses that lie within a cylinder of radius $r_E(z_{\ell}, z_s)$ and thickness $cdt_{\ell}$. It is given by :
\begin{equation}
	\frac{d\tau}{dM} = c \, dt_{\ell} \, \pi r_E^2(z_{\ell}, z_s) \, \frac{d n}{dM}(t_{\ell})
\end{equation}
Here, $\frac{d n}{dM}(t_{\ell})$ is the proper number density of the lenses at lookback time $t_{\ell}$, for lenses with masses that lie between $M$ and $M + dM$.  It is convenient to re-express the differential optical depth exclusively in terms of redshifts and the velocity dispersion. To do so, we invoke the relation between lookback time and redshift:
\begin{equation}
	dt_{\ell} = \frac{1}{H_0}\frac{dz_{\ell}}{(1 + z_{\ell})E(z_{\ell})}
\end{equation}
We further introduce an integrated (or marginalized) number density $n(z_{\ell}) = \int_{M_{\mathrm{min}}}^{M_{\mathrm{max}}} \, \frac{dn}{dM}(z_{\ell}) \, dM$,  where $M_{\mathrm{min}} = 10^{10} M_{\odot}$, $M_{\mathrm{max}} = 10^{15} M_{\odot}$~\footnote{We have chosen this mass range as most of the current halo mass function models are valid inside this mass range. In the actual analysis, this mass range has to be made sufficiently wide so that further extension of the mass range will have no effect on the number of lensed events and the time delay distribution.}. We then define an integrated comoving number density as $n^c(z_{\ell}) \equiv a^3(z_{\ell})n(z_{\ell})$ where $a$ is the scale factor, and a normalized redshift-dependent number density (or, equivalently, a mass distribution with shape-parameter $z_{\ell}$) as $p_M(M; z_{\ell}) = n(z_{\ell}, M)/n(z_{\ell})$. The differential optical depth may now be written as:
\begin{equation}\label{eq:diff-opt-depth}
	\frac{d\tau}{dz_{\ell}d\sigma} = \frac{16\pi^3}{E(z_{\ell})}\frac{c}{H_0}\left(\frac{\sigma}{c}\right)^4 \left[D_{\Delta t}(z_{\ell}, z_s)\right]^2 \left(\frac{D_{\ell s}}{D_s}\right)^4 p_{\sigma}(\sigma; z_{\ell})n^c(z_{\ell})
\end{equation}
where $p_{\sigma}(\sigma; z_{\ell}) \equiv p_M(M;z_{\ell})\frac{dM}{d\sigma}$ is a velocity dispersion distribution at redshift $z_{\ell}$. This differential optical depth is used in Eq.\eqref{eq:P_sigma_zl} to choose the distribution of $\zl$ and $\sigma$.   

\begin{figure}[tbh]
	\includegraphics[scale=0.425]{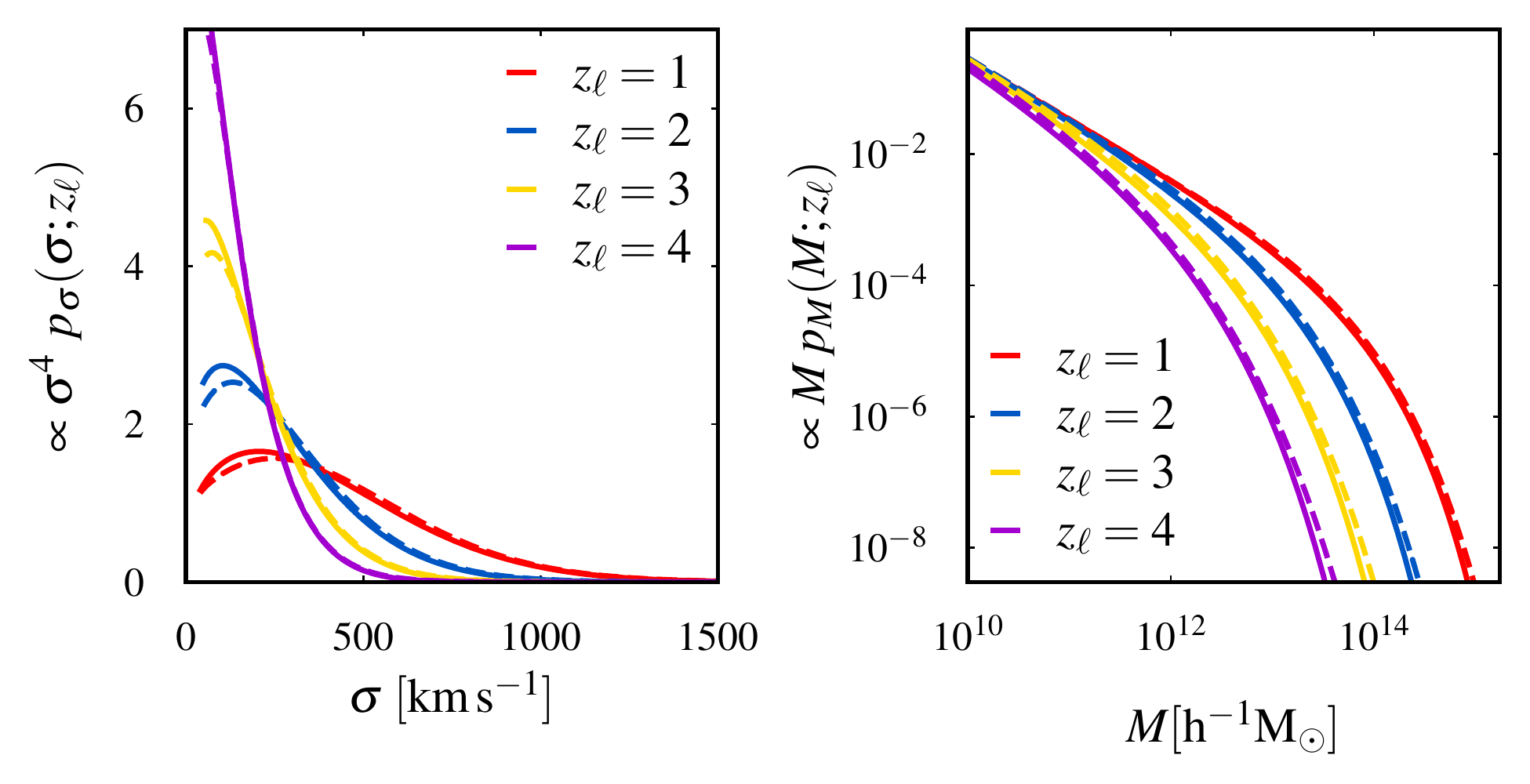}
	\caption{The left panel shows the distribution of lens velocity dispersions along the line of sight evaluated at different redshifts for two different halo mass models, ``Behroozi'' \cite{behroozi2013} (solid lines) and ``Jenkins'' \cite{jenkins2001} (dashed lines). The right panel shows the corresponding mass functions.}
	\label{fig:lens-distribution}
\end{figure}

The distribution of the lens velocity dispersion $p_{\sigma}(\sigma; z_{\ell})$ at different lens redshifts are calculated assuming a halo-mass function and cosmology $\vec{\Omega}$. For illustration, Fig.~\ref{fig:lens-distribution} plots velocity-dispersion distributions for two different halo mass function models at  different redshifts. Notice the non-trivial difference between the distributions pertaining to the different halo mass models. Such differences could potentially lead to biases in the estimation of $\vec{\Omega}$, if the true halo-mass model is not known. See the next section for an illustration of such systematic errors. 

In order to evaluate the differential optical depth, a mapping between the halo mass $M$ and the corresponding dispersion $\sigma$, must be constructed. Assuming spherically symmetric halos with uniform density $\rho$ and radius $R$, that have virialised, we have:
\begin{equation}
	\sigma \simeq \sqrt{\frac{GM}{R}}, ~~ M = \frac{4}{3}\pi R^3 \rho, ~~\rho = \Delta_{\mathrm m}(z)\rho_{\mathrm m}(z),
\end{equation}
where $\Delta_{\mathrm m}$ is the overdensity of the halo and $\rho_{\mathrm m}$ is the mean density of the universe, both $\Delta_{\mathrm m}$ and $\rho_{\mathrm m}$ depends on redshift. 
Eliminating $R$ to express $\sigma$ in terms of $M, \rho$, and differentiating with respect to $M$, yields $dM/d\sigma = 3M/\sigma$. We also need to use some minimum and maximum cutoff for $\sigma$ to compute the total optical depth defined in Eq.\eqref{tau-defined}. The natural choices are $\sigma_\mathrm{min} = \sigma(M_\mathrm{min})$ and $\sigma_\mathrm{max} = \sigma(M_\mathrm{max})$. 

\section{Systematic errors}
\subsection{Due to incorrect halo mass function models}

\begin{figure}[tbh]
	\includegraphics[width=\columnwidth]{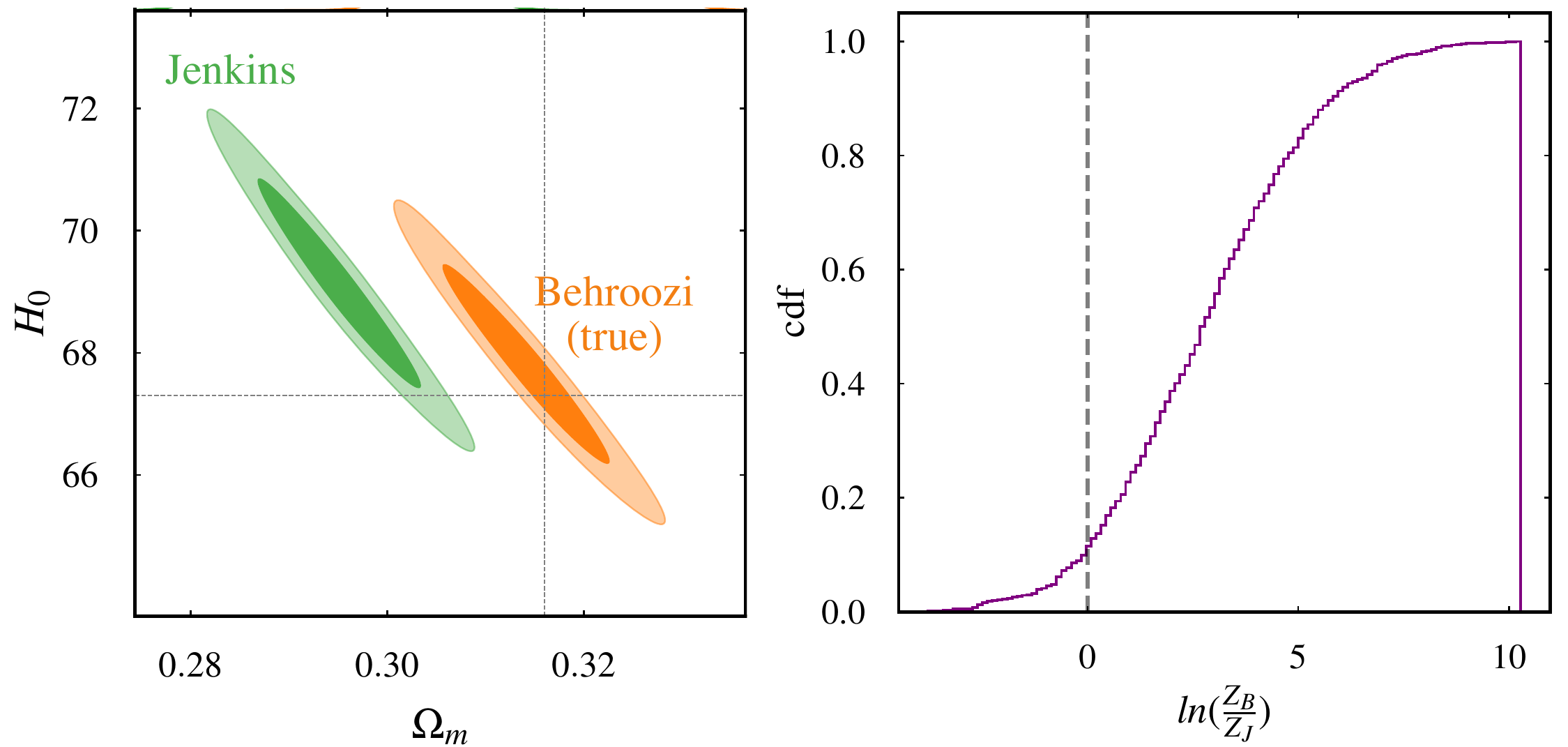}
	\caption{\textit{Left:} The bias in the expected posterior distributions of $H_0$  and $\Omega_m$ due to using an incorrect halo mass function model. The simulated observational data are created assuming the Behroozi model. The orange contours correspond to the posterior estimated assuming the Behroozi (true) model (same as Fig.~2, in the main text), 
		while the green contours correspond to the posteriors estimated assuming the Jenkins model. Using the wrong halo-mass model does not affect the precision, but introduces a systematic bias in the parameter estimation. \textit{Right:} The cumulative distribution of the Bayes factor (likelihood ratio) between the correct and incorrect halo mass function models from a catalog of simulated observation data set. We compute each observational data set ($N$ and $\{\Delta t_i\}$) assuming the  Behroozi model and compute the Bayes factor between the  Behroozi and Jenkins models. It can be seen that the correct model is preferred (Bayes factor $> 1$) for $90\%$ of the simulated catalogs.}
	\label{fig:lensing-hmf}
\end{figure}

A major source of systematic error in this method is the model of the halo-mass-function. We therefore also study the effect of using the wrong halo-mass model to construct the template time delay distributions. In particular, we continue to assume the model of Behroozi \cite{behroozi2013} to be the ``true'' model, but construct the templates using the model by Jenkins \cite{jenkins2001}, implemented in the \textsc{hmfcalc} package \cite{hmfcalc}. For $R = 5\times10^5\;\mathrm{yr}^{-1}$ and $T_{\mathrm{obs}}=10\;\mathrm{yrs}$, we find that, while the precision of the estimates are similar to the one when the templates also used the ``true'' halo mass model, the estimates of the cosmological parameters are significantly biased, and the ``true'' cosmology is not recovered at $68\%$ confidence (Fig.~\ref{fig:lensing-hmf}).  This underlines the need of accurate theoretical models of the distribution of lens properties.

If the data are sufficiently informative, one could also try to mitigate this problem by performing Bayesian model selection using different halo mass function models. If one of the models among a suite of constructed models happens to correspond to the true model, then this should produce the largest Bayesian evidence, denoted by $Z$ in Eq.~(0.3), in the main text. 
We perform a simple illustration of this by considering two models: Behroozi   \cite{behroozi2013} and Jenkins \cite{jenkins2001}. Assuming Behroozi to be the true model, and evaluating the pairwise Bayes factor (ratio of evidences) for Jenkins model, we find that Behroozi model is consistently selected as the most favored model across multiple realisations of the set of detected lensed events (Fig.~\ref{fig:lensing-hmf}). This also will provide a means to constrain the halo mass function models using observations. A more robust alternative would be to jointly constrain the parameters of the halo-mass model with the cosmological parameters, which we leave for future work.

\subsection{Due to incorrect identification of lensed events}

In our analysis we have assumed that lensed pairs are identified with complete certainty, which is not true in reality. Any identification method will have a tunable false positive probability $\alpha$  (fraction of falsely identified lens pairs) and a true positive probability $\epsilon$ (fraction of truly identified lensed pairs). For a given $\alpha$ and $\epsilon$, it is easy to show that contamination fraction is $\kappa \sim \alpha N / \epsilon \tau$, where $N$ is the total number of events and $\tau$  is the lensing probability. Assuming $N \sim 10^6$ and $\tau \sim 1\%$, if we want the contamination fraction to be small ($\sim10\%$), this require our lensing identification method to have a true positive rate of  $\epsilon \sim 0.5$ for a false positive rate of $\alpha \sim 10^{-9}$ per pair. The current (sub-optimal) methods are able to achieve $\epsilon \sim 0.5$ for $\alpha \sim 10^{-8}$ in current generation detectors (by extrapolating Fig.~9 of \cite{haris2018}). These sub-optimal methods could be further improved by including more information of the source and lens properties, such as the Morse phase, magnification ratio, etc~\cite{Janquart:2022wxc,Janquart:2023osz,More:2021kpb}. Additionally, In third generation detectors, we will have better ability distinguish between lensed and unlensed events (e.g., due to the improved precision in estimating source parameters and by measuring additional signal properties such as spin angles and higher harmonics). Thus, we expect that it will be possible to keep the contamination fraction sufficiently small. 

In addition, even this contamination could be modelled and accounted for in the analysis. Suppose we set the threshold such that there is a (low) expected contamination fraction  (for some fiducial lensing rate). Then we can account for the contamination by constructing template time delay distributions as 
\begin{equation}
	p\left(\Delta t~|~\vec{\Omega}\right) = \frac{\kappa}{1+\kappa} ~ p_\mathrm{unlensed} (\Delta t) + \frac{1}{1+\kappa} p_\mathrm{lensed} (\Delta t~|~\vec{\Omega})
\end{equation}
In reality the relation between $\alpha$  and $\epsilon$  will be a function of source parameters, such as redshift, so we have to derive their relationship through simulations. After determining $\alpha(\epsilon)$  we can still recover cosmological information. 

In summary, we can control the expected fraction of contaminants and they will have a known time delay distribution (which differs  from that of the lensed events; see, e.g, Fig 2 of ~\cite{haris2018}), so we can use this to account for their effect without significantly losing our constraining ability. We will study this in detail in future work. 


\end{document}